# Blood-coated sensor for high-throughput ptychographic cytometry on a Blu-ray disc

Shaowei Jiang[1, 2], Chengfei Guo[1, 2*], Tianbo Wang[1, 2], Jia Liu[1], Pengming Song[1], Terrance Zhang[1], Ruihai Wang[1], Bin Feng[1], and Guoan Zheng[1,*]

[1]Department of Biomedical Engineering, University of Connecticut, Storrs, CT 06269, USA
[2]These authors contributed equally to this work.

*Correspondence:
C. G. (chengfei.guo@uconn.edu) or G. Z. (guoan.zheng@uconn.edu)

**Abstract**
Blu-ray drive is an engineering masterpiece that integrates disc rotation, pickup head translation, and three lasers in a compact and portable format. Here we integrate a blood-coated image sensor with a modified Blu-ray drive for high-throughput cytometric analysis of various bio-specimens. In this device, samples are mounted on the rotating Blu-ray disc and illuminated by the built-in lasers from the pickup head. The resulting coherent diffraction patterns are then recorded by the blood-coated image sensor. The rich spatial features of the blood-cell monolayer help down-modulate the object information for sensor detection, thus forming a high-resolution computational bio-lens with a theoretically unlimited field of view. With the acquired data, we develop a lensless coherent diffraction imaging modality termed rotational ptychography for image reconstruction. We show that our device can resolve the 435 nm linewidth on the resolution target and has a field of view only limited by the size of the Blu-ray disc. To demonstrate its applications, we perform high-throughput urinalysis by locating disease-related calcium oxalate crystals over the entire microscope slide. We also quantify different types of cells on a blood smear with an acquisition speed of ~10,000 cells per second. For *in vitro* experiment, we monitor live bacterial cultures over the entire Petri dish with single-cell resolution. Using biological cells as a computational lens could enable new intriguing imaging devices for point-of-care diagnostics. Modifying a Blu-ray drive with the blood-coated sensor further allows the spread of high-throughput optical microscopy from well-equipped laboratories to citizen scientists worldwide.

Traditional light microscope has broad applications in fields related to physical and life sciences. The intrinsic compromise between imaging resolution and field of view, however, imposes a limit on the system throughput (1-3). One can have a large field of view with poor resolution, or a small field of view with good resolution, but not both. The information transmitted by the microscope can be quantified using the concept of space-bandwidth product (SBP), which characterizes the total number of effective independent pixels of a system's field of view (1). The SBP of most off-the-shelf objective lenses is on the order of ~10 megapixels, regardless of their magnification factors or numerical apertures (NAs). For example, a regular 20×, 0.4 NA objective lens has a resolution of 0.8 µm and a field of view of ~1 mm in diameter, corresponding to an SBP of ~6 megapixels.

In the past years, different imaging strategies have been developed to increase the SBP of a microscope platform (2). For lens-based systems, one example is Fourier ptychographic microscopy (FPM) (4), which computationally synthesizes the captured data into a high-SBP image in the Fourier domain. However, the two-dimensional (2D) aperture synthesizing process in FPM cannot be directly applied for three-dimensional (3D) objects (3, 5, 6). Therefore, it is challenging for FPM to image thicker bio-specimens such as thick tissue sections or bacteria colonies on thick agar plates. Another strategy to address the tradeoff between resolution and field of view is to adopt a giant objective lens for data acquisition (7, 8). However, the required number of optical surfaces of the lens results in an expensive, bulky, and highly specialized design with a tightly controlled usage range. They are not readily accessible to most research laboratories due to the associated costs and maintenance issues.

Lensless coherent diffraction imaging (CDI) offers an alternative for high-SBP microscopy without using any physical lens (9). In a typical implementation, a spatially coherent light source is used to illuminate the object. The resultant coherent diffraction patterns are acquired using a 2D image sensor downstream. Instead of capturing a direct image of the object, the sensor registers the squared amplitude of the scattered waves at the detector plane. The phase information that characterizes the optical delay accrued during propagation is lost in this acquisition process, thereby preventing the reconstruction of the real-space object through back propagation. For this reason, lensless CDI requires the retrieval of the missing phase distribution at the detector plane, which can be achieved using an iterative reconstruction process known as phase retrieval (10).

Here, we report the development of a lensless cytometric analysis system using a Blu-ray drive and a blood-coated image sensor. Blu-ray drive is an engineering masterpiece that integrates disc rotation, optical pickup head translation, and three lasers in a compact and portable format. In our device, we mount different bio-specimens on the rotating Blu-ray disc and illuminate them using the built-in lasers from the Blu-ray drive. The optical pick-up head mounted on the translational stage in the Blu-ray drive is replaced with a blood-coated image sensor for lensless diffraction data acquisition. The dense monolayer of blood cells on the sensor chip can down-modulate the object information for detection, thus forming a high-resolution computational bio-lens with a theoretically unlimited field of view. In contrast to the Blu-ray drive's inherently rapid rotation, we rotate the disc at a speed of <0.1 degrees per second. This extremely low speed allows for the continuous and efficacious acquisition of the corresponding coherent diffraction patterns without motion blurs. It also makes our device compatible with most bio-specimens, including cell cultures on Petri dishes.

At the heart of our imaging process is a new lensless CDI modality termed rotational ptychography. This new approach models the disc spinning process and recovers the high-resolution object wavefront with both intensity and phase information. With this approach, the reported device can resolve the 435 nm linewidth on the resolution target. Complex object wavefronts over the entire blood-coated image sensor (~30 mm$^2$) can be acquired in 15 seconds and wide field-of-view images with gigapixel SBP can be obtained in minutes. The achieved image acquisition throughput is comparable to that of high-end whole slide scanners (11). The final field of view is only limited by the size of the Blu-ray disc and we image the entire Petri dish for tracking the growth of bacterial colonies on a thick agar plate. The recovered complex wavefront can be further propagated to different axial positions for 3D refocusing.

To the best of our knowledge, it is the first demonstration of using a blood-cell monolayer as a high-SBP bio-lens for lensless CDI. Since one can readily obtain blood from finger pricks, the reported scheme would enable new intriguing and turnkey devices for high-throughput microscopy. Miniaturization of existing automated microscope to produce inexpensive portable devices that duplicate the full range of laboratory-based high-throughput imaging capability can benefit the fields of microbiology, hematology, telemedicine, urine sediment examination, and various point-of-care diagnostics applications. The modification of optical disc drive for gigapixel microscopy can also increase the accessibility of high-throughput cytometric analysis from well-equipped laboratories to citizen scientists worldwide.

**Materials and methods**

**Hacking a Blu-ray drive for coherent diffraction imaging.** A typical Blu-ray drive is designed to be backward compatible with both digital video disc (DVD) and compact disc (CD). As a result, it contains three lasers in the optical pickup head, with a 405 nm ultraviolet laser for the Blu-ray disc, a 650 nm red laser for the DVD, and a 780 nm near-infrared laser for the CD (12). Similar to the operation of DVD and CD drives, Blu-ray drive functions by focusing the laser beam from the optical pickup head into a diffraction-limited spot on the data pits of the disc, where the pit depth is one-fourth of the laser wavelength (13). When this laser spot hits a pit, the reflection is destructively interfered upon, and the pickup head receives a resulting signal of '0'. Conversely, if no pit is hit, the higher intensity reflection results in a signal of '1'.

Figure 1a shows the schematic of the reported device, where we modified a commercially available Blu-ray drive (LG WH14NS40, Amazon) for high-throughput CDI. A key component for CDI is the coherent light source. In some of the previous lensless microscopy demonstrations, LEDs were used as the light source for sample illumination (14-18). A major disadvantage of the LED source, however, is the low spatial and temporal coherence that deteriorates the reconstruction quality for the phase retrieval process. In addition, the low light intensity of LEDs requires a long exposure time for data acquisition (an exposure time of >0.2 seconds is often required after proper spatial and temporal filtering). As a result, the sample must remain stationary during the exposure phase of each acquisition. In our implementation, we couple the built-in 405-nm and 650-nm lasers from the Blu-ray drive to optical fibers for sample illumination. The high illumination intensity from the laser source results in an exposure time of <4 ms for image acquisition. Therefore, the user can afford to place the sample on the continuous spinning disc during the acquisition process, enabling high-throughput CDI without complications arising from motion blur. In Figure S1a, we provide the detailed wiring diagrams of the two laser diodes in the Blu-ray drive.

As shown in Figure 1a, different bio-specimens are mounted on the spinning disc of the drive. Here, the optical pick-up head on the translation stage is replaced with a blood-coated image sensor for diffraction pattern acquisition. Figures S1b shows the wiring diagram of the rotary motor of the spinning disc. Figure S1c shows the wiring diagram of translation motor for the blood-coated image sensor. In our device, we used an Arduino micro-controller to program the motions of these two motors. The operation of the entire platform can be found in Supporting Information, Video 1, where we demonstrate that the combination of disc rotation and sensor translation enabled us to utilize the total area of the Blu-ray disc as the imaging field of view.

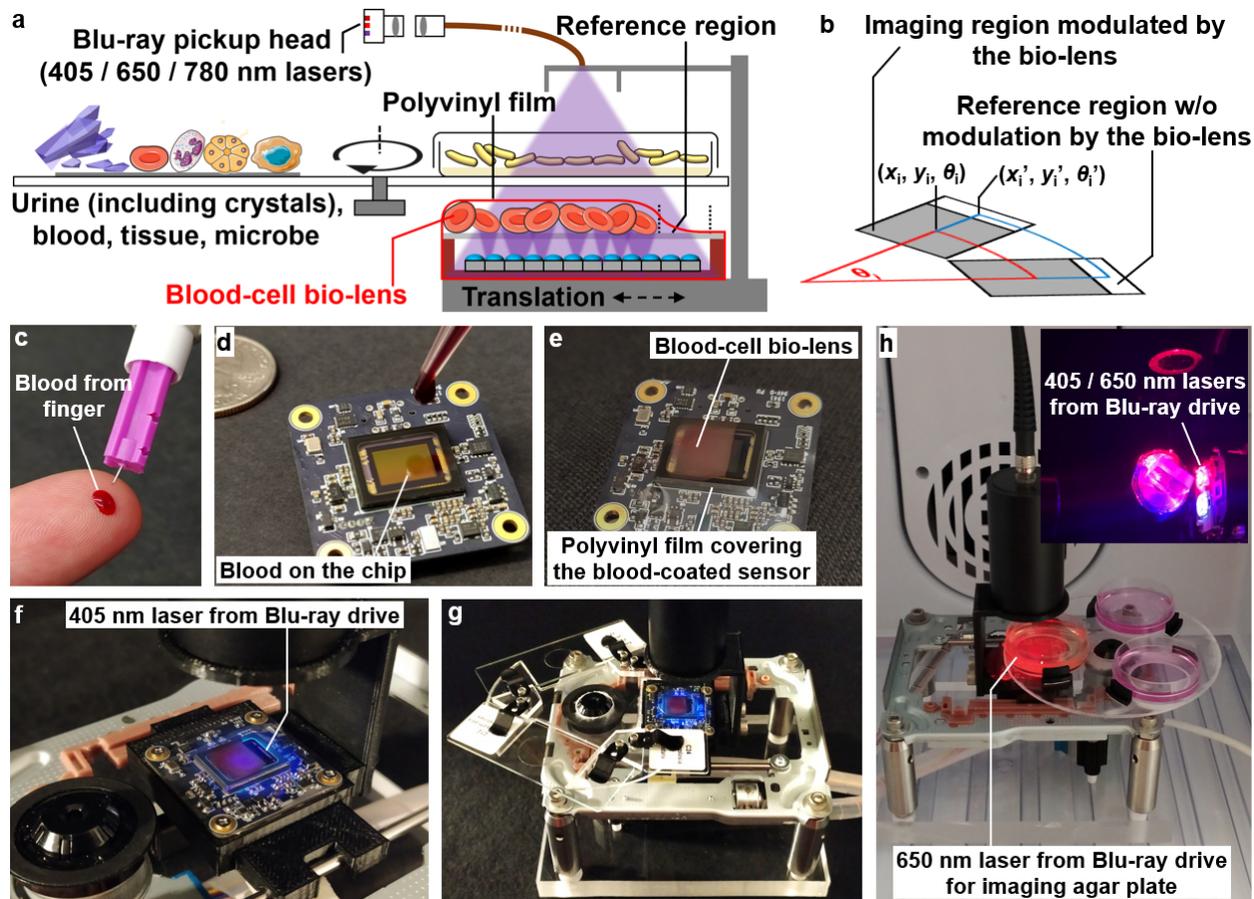

**Fig. 1. Hacking a Blu-ray drive for high-throughput CDI.** (a) Schematics of the reported device, where the lasers from the optical pickup head are coupled to optical fibers for sample illumination. A blood-coated image sensor is placed under the samples for image acquisition. A polyvinyl film is stretched over the blood-coated sensor chip, preventing direct contact between the blood cells and external objects. (b) The rotational ptychographic reconstruction routine for the reported device. For the $i^{th}$ captured image, the reference region without blood cells is used to track the positional shift $(x'_i, y'_i)$ and rotation angle $\theta'_i$ of the sample. The positional shift and rotation angle of the imaging region $(x_i, y_i, \theta_i)$ can then be inferred from $(x'_i, y'_i, \theta'_i)$. (c) Finger prick for obtaining blood samples. (d) The blood is smeared on top of the image sensor and fixed with alcohol. (e) The blood-coated chip is covered with a stretched plastic wrap for protection. (f) The blood-coated sensor is mounted on the translation stage of the Blu-ray drive. (g) Bio-specimens are mounted on the rotating disc for CDI. (h) The device can be placed in an incubator for monitoring live cell cultures. Inset shows the 650 nm and 405 nm laser beams from the built-in lasers of the Blu-ray drive. Refer to Supporting Information, Video 1 for a demonstration of its operation.

The schematic in Figure 1a also shares similarities with the concept of lab on a disc. In early demonstrations, the self-referencing interferometry concept of the data pit was adopted for biosensing (13, 19). The disc spinning process can also be used for liquid sample loading via centrifugal force (20, 21). In our device, the rotational motion is not used for liquid sample loading. Instead, spinning the sample directly on top of the blood-coated image sensor introduces rotational diversity measurements for CDI in Figure 1b.

**Computational blood-cell bio-lens.** In previous demonstrations, it has been shown that individual biological cells or microorganisms can serve as physical microlens for sub-diffractive focusing and microscopy imaging (22, 23). In the reported device, we employ a dense monolayer of blood cells as a computational bio-lens with a theoretically unlimited field of view, thereby addressing the trade-off between resolution and imaging area of a conventional objective lens. The operation of this blood-cell monolayer is similar to that of structured illumination microscopy, where a non-uniform illumination pattern is used to down-modulate the high-frequency object information for detection. In our implementation, the non-uniform illumination pattern is replaced by a monolayer blood-cell at the detection path. In contrast with illumination-based methods (4, 15, 24, 25), the recovered image of our device depends solely on how the complex wavefront exits the sample (26, 27). Therefore, the sample thickness becomes irrelevant for the imaging model. We can digitally propagate the recovered wavefront to any position along the axial direction for 3D refocusing.

As shown in Figure 1c, we followed the blood sugar test protocol to obtain blood from finger prick. We then smeared it on top of an image sensor and fixed it with ethyl alcohol in Figure 1d. Figure 1e shows a method to protect the blood-cell monolayer from external contacts by covering it with a thin polyvinyl film (Kirkland plastic wrap, Costco). This protective film also addresses potential bio-safety concerns related to contacts with human whole blood. However, a contaminated lancet or an improper procedure of drawing blood would spread bloodborne diseases. Low-risk alternatives include fish blood from supermarket and blood phantoms. We note that, disorder-engineered metasurface can also serve as the large-scale scattering lens in our device. Such surfaces are free of ethical and contamination issues. However, the fabrication process often involves photolithography and other advanced material deposition / etching techniques. For example, phase modulation of the metasurface can be achieved via chemical vapor deposition followed by lithographic patterning and etching (28).

The advantages of using blood-cell monolayer as computational scattering lens can be summarized as follows. First, smearing the blood on the sensor's coverglass is simple and requires no sophisticated tools. In contrast, fabrication of metasurface often involves photolithography, e-beam lithography, or other advanced material processing techniques. Second, the rich spatial feature of the blood cells can effectively modulate the incoming light waves for detection. The blood smearing process allows for the formation of a monolayer of cells on the sensor chip. Light interaction with this monolayer can be modeled by a simple multiplication between the light waves and the complex profile of the monolayer. In contrast, conventional disordered media often require the measurement of the full transmission matrix (29, 30). The resulting heavy demand for computer memory prevents its applications in wide-field CDI. Third, the blood-cell layer enables both intensity and phase modulation of the incoming light waves. It allows the quantitative recovery of object phase information regardless of the spatial frequency contents. In a later section, we demonstrate imaging various types of crystals in urine samples for rapid sediment examination. The phase images of these crystals contain slow-varying information with many $2\pi$ wraps, which are difficult to obtain using other common CDI techniques (14, 27, 31, 32).

Our prototype device is shown in Figures 1f-1g, where the blood-coated sensor is mounted on the translation stage and the specimens are mounted on the rotating disc. Furthermore, Figure 1h shows that the entire device can be placed in an incubator for imaging live cell cultures on the Petri dishes. The inset of Figure 1h shows the 405 nm and 650 nm built-in lasers from the optical pickup head of the Blu-ray drive. For imaging bacteria cultures, we used the 650 nm laser as the light source. For other experiments, we used the 405 nm laser as the light source.

**Rotational ptychography.** Robust phase retrieval for CDI is typically performed by introducing different diversity measurements to constrain the solution space (33, 34). Typical diversity strategies include multi-height (35-37), multi-wavelength (31, 32, 38), and transverse translation measurements (25, 27, 34, 39-48). The strategy of using transverse translation diversity measurements for CDI is also termed ptychography, a technique originally proposed for electron microscopy (49) and then brought to its modern form using the iterative phase retrieval framework (39). In a typical implementation, ptychography illuminates the sample with a spatially confined probe beam. The specimen is then laterally translated through the confined beam and the diffraction patterns are recorded at the reciprocal space. The phase retrieval process iterates between two domains: real space and reciprocal space. In the real space, the confined probe beam limits the physical extent of the object for each measurement and serves as a compact support constraint. In the reciprocal space, the diffraction measurements serve as the Fourier magnitude constraints.

Here, we report a new type of diversity measurement based on the disc spinning process of the optical drive. In contrast with the conventional ptychography, we term it rotational ptychography to highlight the rotational motion during the disc spinning process. As shown in Figures 1a and 1b, we divided the captured images into two regions: a reference region without modulation by the blood-cell monolayer, and an imaging region with blood-cell modulation. For the reference region, we use $(x_i', y_i')$ to denote the x-y positional shift between the first and the $i^{th}$ captured image. We use $\theta_i'$ to denote the rotation angle of the $i^{th}$ captured image, with the rotation axis at the center of this region. Similarly, we can define the positional shift $(x_i, y_i)$ and rotation angle $\theta_i$ for the imaging region.

Direct tracking of the disc motion $(x_i, y_i, \theta_i)$ using the imaging region is challenging because the diffracted light waves have been encoded by the blood-cell monolayer profile. In conventional ptychography, the transverse translational motion is identical for the entire image. Therefore, one can use the reference region for positional tracking and the results can be applied to the modulation region of the same image (45, 48). In the reported rotational ptychography, the disc motion is different at different regions of the captured images, i.e., $x_i \neq x_i', y_i \neq y_i'$ in Figure 1b. As a result, we cannot use the tracking results from the reference region to infer the motion at the modulation region.

To address this challenge, we developed a 3-step procedure to recover the disc motion of our device. In step 1, we adopt an incremental tracking strategy where we assume that the rotation is negligible in-between adjacent captured images. The positional shift $(\Delta x_j', \Delta y_j')$ in-between the $j^{th}$ and $(j-1)^{th}$ captured images at the reference region can be calculated via cross-correlation analysis (50). The positional shift between the first captured image and the $i^{th}$ captured image $(x_i', y_i')$ can then be approximated by the summation of these incremental shifts: $\left(x_i' = \sum_{j=1}^{i} \Delta x_j', y_i' = \sum_{j=1}^{i} \Delta y_j'\right)$. This incremental tracking strategy allows us to recover a good initial estimation of $(x_i', y_i')$ for subsequent processing. In comparison, Figure S2 shows result of the direct recovery of $(x_i', y_i')$ via cross-correlation analysis, where the rotation angle prevents the correct recovery of the positional shift. In step 2, we refine $(x_i', y_i')$ and estimate the rotation angle $\theta_i'$ at the reference region via a close-form equation detailed in Supplementary Note 1. This process is only effective with a good initial estimate of $(x_i', y_i')$ from step 1. Figure S3 shows the failed recovery of $(x_i', y_i', \theta_i')$ without implementing step 1. In step 3, we obtain the rotation angle of the imaging region by setting $\theta_i = \theta_i'$. We then rotate the entire captured images by an angle of $-\theta_i$, with the rotation axis centered at the imaging region. After this operation, there is no rotation in-between the process images. We can then obtain the positional shift at the imaging region $(x_i, y_i)$ by performing cross-correlation analysis of the reference region (50). With the recovered disc motion $(x_i, y_i, \theta_i)$ of the spinning process, we then perform ptychographic reconstruction based on all acquisitions.

## Rotational ptychography – a new type of diversity measurement based on the disc spinning process

**Input:** Captured raw images $I_i$ ($i = 1, 2, \cdots, I$), the blood-cell bio-lens profile $B(x,y)$, the estimated spinning angles $\theta_i$ of the disc, and positional shifts $(x_i, y_i)$
**Output:** High-resolution object $O(x,y)$

1. Initialize object wavefront $W(x,y)$ on the blood-cell bio-lens plane
2. **for** $n = 1: N$ (different iterations)
3.     **for** $i = 1: I$ (different measurements)
4.         $W_i^x(x', y') = \mathcal{F}_x^{-1}\left\{\exp\left(-2\pi i u \tan\frac{-\theta_i}{2} y\right) \mathcal{F}_x\{W(x,y)\}\right\}$    % $\mathcal{F}_x$: 1D FT along x-direction
5.         $W_i^{yx}(x', y') = \mathcal{F}_y^{-1}\left\{\exp(-2\pi i v (\sin\theta_i) x) \mathcal{F}_y\{W_i^x(x,y)\}\right\}$    % $\mathcal{F}_y$: 1D FT along y-direction
6.         $W_i^{xyx}(x', y') = \mathcal{F}_x^{-1}\left\{\exp\left(-2\pi i u \tan\frac{-\theta_i}{2} y\right) \mathcal{F}_x\{W_i^{yx}(x,y)\}\right\}$ % $W_i^{xyx}$: wavefront after rotation of $\theta_i$
7.         $W_i^{shift}(x, y) = W_i^{xyx}(x - x_i, y - y_i)$    % Shift the wavefront based on the positional shifts
8.         $\varphi_i(x, y) = B(x, y) \cdot W_i^{shift}(x, y)$    % Exit wave on the bio-lens plane
9.         $\phi_i(k_x, k_y) = \mathcal{F}(\varphi_i(x, y)) \cdot H_{free}(d_0)$    % Propagate to the sensor plane
10.       $\Psi_i(k_x, k_y) = \phi_i(k_x, k_y) \cdot CTF_{angular}(k_x, k_y)$ % Coherent angular transfer function of the pixels
11.       $\psi_i(x, y) = \mathcal{F}^{-1}\left(\Psi_j(k_x, k_y)\right)$    % Light on the image sensor plane
12.       Update $\psi_i(x, y)$:

          $\psi_i'(x, y) = \psi_i(x, y) \cdot \dfrac{\sqrt{I_i(\lceil x/M \rceil, \lceil y/M \rceil)}}{\sqrt{U_i(\lceil x/M \rceil, \lceil y/M \rceil)}}$    % '$\lceil\ \rceil$' presents the ceiling function
                                                                           % $U_i(\lceil x/M \rceil, \lceil y/M \rceil) = |\psi_i(x,y)|^2 * ones(M,M)_{\downarrow M}$

13.       $\Psi_j'(k_x, k_y) = \mathcal{F}(\psi_j'(x, y))$
14.       Update the exit wave in the Fourier domain using the ePIE algorithm:

          $\phi_j'(k_x, k_y) = \phi_j(k_x, k_y) + \dfrac{conj\left(CTF_{angular}(k_x,k_y)\right) \cdot \left(\Psi_j'(k_x,k_y) - \Psi_j(k_x,k_y)\right)}{|CTF_{angular}(k_x,k_y)|_{max}^2}$

15.       $\varphi_j'(x, y) = \mathcal{F}^{-1}(\phi_j'(k_x, k_y) \cdot H_{free}(-d))$    % Propagate back to the bio-lens plane
16.       Update the shifted object wavefront $W_i^{shift}(x, y)$ using the rPIE algorithm:

          $W_i^{shift'}(x, y) = W_i^{shift}(x, y) + \dfrac{conj(B(x,y)) \cdot (\varphi_i'(x,y) - \varphi_i(x,y))}{(1 - \alpha_w)|B(x,y)|^2 + \alpha_w|B(x,y)|_{max}^2}$

17.       $W_i^{xyx'} = W_i^{shift'}(x + x_i, y + y_i)$    % Shift the updated wavefront back
18.       $W_i^{yx'}(x', y') = \mathcal{F}_x^{-1}\left\{\exp\left(-2\pi i u \tan\frac{\theta_i}{2} y\right) \mathcal{F}_x\{W_i^{xyx'}(x,y)\}\right\}$ % Rotate back the wavefront
19.       $W_i^{x'}(x', y') = \mathcal{F}_y^{-1}\left\{\exp(-2\pi i v (-\sin\theta_i) x) \mathcal{F}_y\{W_i^{yx'}(x,y)\}\right\}$ % Can be combined with lines 4-6
20.       $W(x', y') = \mathcal{F}_x^{-1}\left\{\exp\left(-2\pi i u \tan\frac{\theta_i}{2} y\right) \mathcal{F}_x\{W_i^{x'}(x,y)\}\right\}$
21.     **end**
22. **end**
23. $O(x, y) = W(x, y) * PSF_{free}(-d_1)$    % Propagate the wavefront back to the object plane

**Fig. 2. The reconstruction process for rotational ptychography.** With the recovered disc motion $(x_i, y_i, \theta_i)$, the captured images $I_i(x, y)$ ($i = 1, 2, 3, \ldots$) are used recover the high-resolution complex object $O(x, y)$. The process to retrieve disc motion $(x_i, y_i, \theta_i)$ can be found in Supporting information, Note 1.

The imaging model of our device can be expressed as

$$I_i(x, y) = \left|W(x\cos\theta_i - y\sin\theta_i - x_i, y\cos\theta_i + x\sin\theta_i - y_i) \cdot B(x,y) * PSF_{free}(d_0) * PSF_{angular} * PSF_{pixel}\right|_{\downarrow M}^2, \quad (1)$$

where $I_i(x, y)$ is the $i^{th}$ captured image, $W(x, y)$ is complex exit wave from the specimen, $B(x, y)$ is the complex modulation profile of the blood-cell monolayer, $PSF_{free}(d_0)$ is the convolution kernel for free-space propagation of distance $d_0$, $PSF_{angular}$ is the convolution kernel for modeling the angular response of the pixels (45), $PSF_{pixel}$ is the convolution kernel for modeling the spatial response of the pixels, '↓ $M$' represents $M$ by $M$ downsampling, '*' denotes convolution, and the term '$x\cos\theta_i - y\sin\theta_i - x_i, y\cos\theta_i + x\sin\theta_i - y_i$' models the motion of the sample on the spinning disc.

The detailed reconstruction routine is shown in Figure 2. The profile of the blood-cell monolayer can be obtained from a calibration experiment and the same profile can be enforced in all subsequent experiments. In line 14 of Figure 2, we update the exit wave in the Fourier domain using the ePIE routine (42). In line 16, we update the exit wave in the spatial domain using the rPIE routine (51). Once we obtain the exit wave $W(x, y)$, we can digitally propagate it back to any position along the axial direction in line 23. Without optimizing the reconstruction speed, the current processing time for 450 raw images with 2048 by 2048 raw pixels each is ~12 mins using a Dell Aurora R12 computer. Because of the Fourier shearing operations in lines 4-6, the computational cost is about 50% higher than that of common lensless phase retrieval process. It is also possible to use parallel programming platforms and streamlined C++ code to substantially shorten the processing time with 10-100 folds.

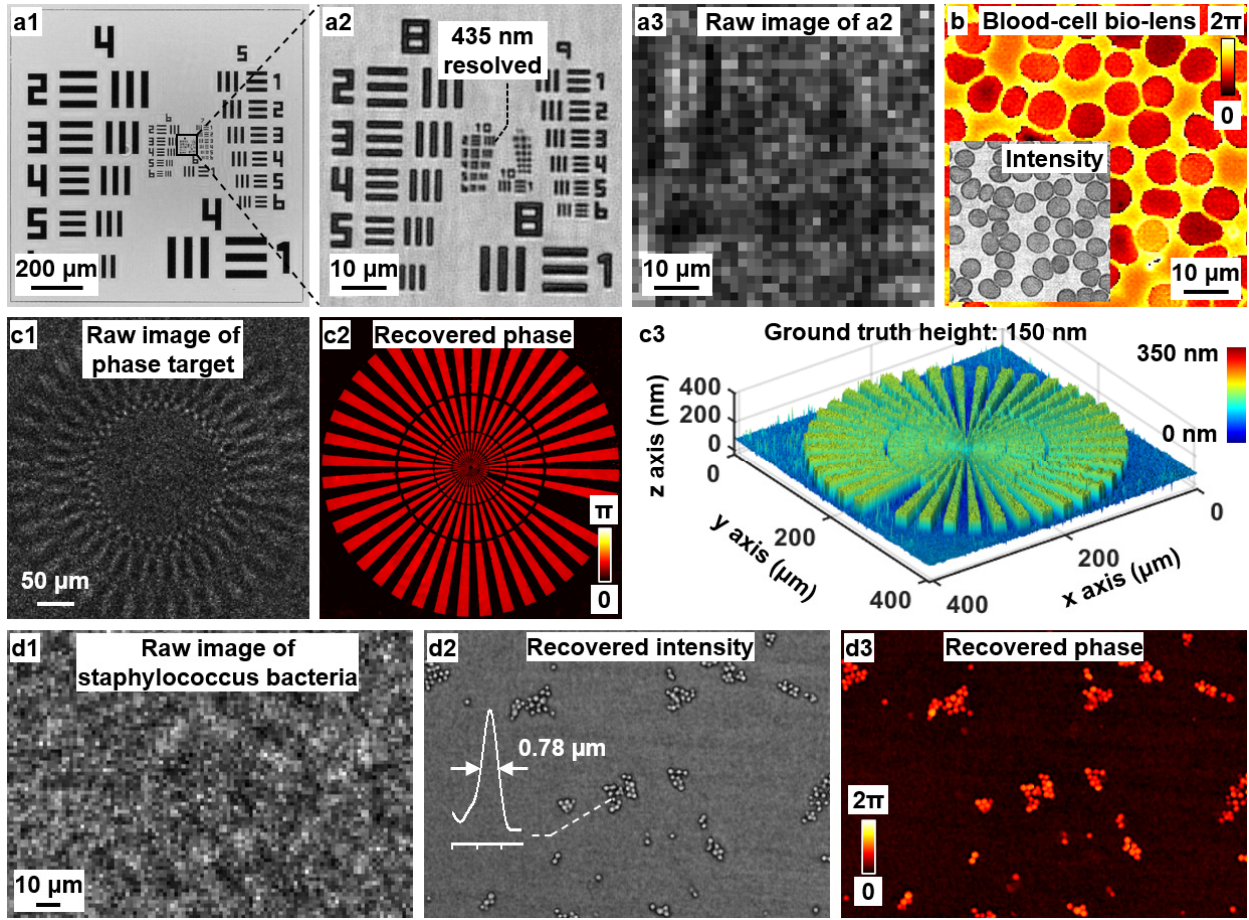

**Fig. 3. Imaging performance characterization using resolution targets.** (a1-a2) The recovered image of a resolution target, where we can resolve the 435 nm linewidth. (a3) The capture raw image corresponding to (a2). (b) The recovered image of blood-cell bio-lens from a calibration experiment. (c1) The captured raw image of a phase target. (c2-c3) The recovered phase and the height map of the phase target, with a ground-truth step height of 150 nm. (d1) The captured raw image of staphylococcus bacteria. (d2-d3) The recovered intensity and phase of the staphylococcus bacteria sample, where we can clearly resolve the individual grape-like bacterial cells.

**Results**

**Imaging performance characterization.** In Figure 3, we validated the imaging performance using a resolution target and a quantitative phase targe. In these experiments, we acquired 450 raw measurements in 15 seconds, with the disc in continuous spinning motion. The acquired images were then used to recover the high-resolution object profiles. Figures 3a1 and 3a2 show the recovered intensity image of a resolution target, where we can resolve the 435 nm linewidth at group 10, element 2. The corresponding raw image of the resolution is shown in Figure 3a3, with a raw pixel size of 1.85 μm and an up-sampling factor $M = 5$. From a calibration experiment, we also recovered the phase and intensity of the blood-cell monolayer smeared on the image sensor, as shown in Figure 3b. We note that the recovered phase of blood cells is lower than that of the background. This is due to the effect of $2\pi$ phase wrapping at the 405 nm laser wavelength. Figure 3c1 shows the raw image of a phase target and Figure 3c2 shows the recovered phase. In Figure 3c3, we plot the recovered height map of the phase target, and the result is in good agreement with the ground-truth height from the manufacturer. In Figure 3d, we further tested the device for imaging small features of staphylococcus bacteria. Under the regular microscope, they appear spherical and form in grape-like clusters. Figure 3d1 shows the captured raw image using the reported device. Figure 3d2 and 3d3 show the recovered phase and intensity images, where we can clearly resolve individual grape-like bacterial cells. A line-trace plot is also provided in Figure 3d2.

**Rapid and quantitative urine sediment examination.** Urinalysis is one of the major *in vitro* diagnostic screening tests in clinical practice. Urine sediment examination offers a direct indication of the state of the renal and genitourinary system. However, traditional microscopy inspection of urine sediments is labor-intensive and time-consuming. The analysis is often based on a small field of view of the urine sediment slide, leading to imprecise results with wide variability. The reported device provides a low-cost and turnkey solution for rapid urine sediment examination. Compared to the small field of view of the traditional light microscope, we can acquire the images of the entire urine sediment slide with a centimeter-scale field of view in ~50 seconds. The acquired images were then used to recover the phase and intensity images of the slide.

Figure 4 shows the results of using the reported device for high-throughput urine sediment examination. Figure 4a shows the recovered phase image of the entire sediment slide. Figure 4b shows the automated tracking of calcium oxalate crystals (labeled by the blue dots) over the entire field of view. Calcium oxalate crystals can be commonly found in healthy urine. However, a large number of calcium oxalate crystals are heavily associated with kidney diseases and can cause some serious health problems. Therefore, the capability of performing automatic and quantitative tracking of calcium oxalate crystals can serve as an indicator for the urine screening test. In our implementation, we first performed image binarization to distinguish the background and different sediments from the recovered phase. The calcium oxalate crystals were then tracked based on the maximum phase after phase unwrapping and the element size.

Based on the recovered images using the reported device, we can perform a quantitative analysis of different elements in the urine samples. In the bar chart of Figure 4c, we analyzed 4 urine samples from different donors and counted the numbers of different elements over the centimeter-scale field of view. These elements include hippuric acid crystals, uric acid crystals, calcium oxalate crystals, triple phosphate crystals, cystine crystals, red blood cells, white blood cells, epithelial cells, and other unidentified sediments. For the automatic tracking of calcium oxalate crystals, our algorithm achieved a high accuracy of 99.5% when compared to the manual count using the regular light microscope. Figure S4 shows several recovered images of irregular crystals that were miscounted as calcium oxalate crystals by the algorithm. A better image classification algorithm or a deep learning approach can be applied to improve the classification results. In Figure 4d-4k, we show the recovered phase images of different representative elements from the urine samples. Figure S5 shows extra images of unidentified sediments, which demonstrate the rich information obtained from the recovered phase image. Figure S6 shows the comparison between the recovered phase images and the intensity images captured by a regular light microscope with a 20×, 0.75 numerical aperture objective lens.

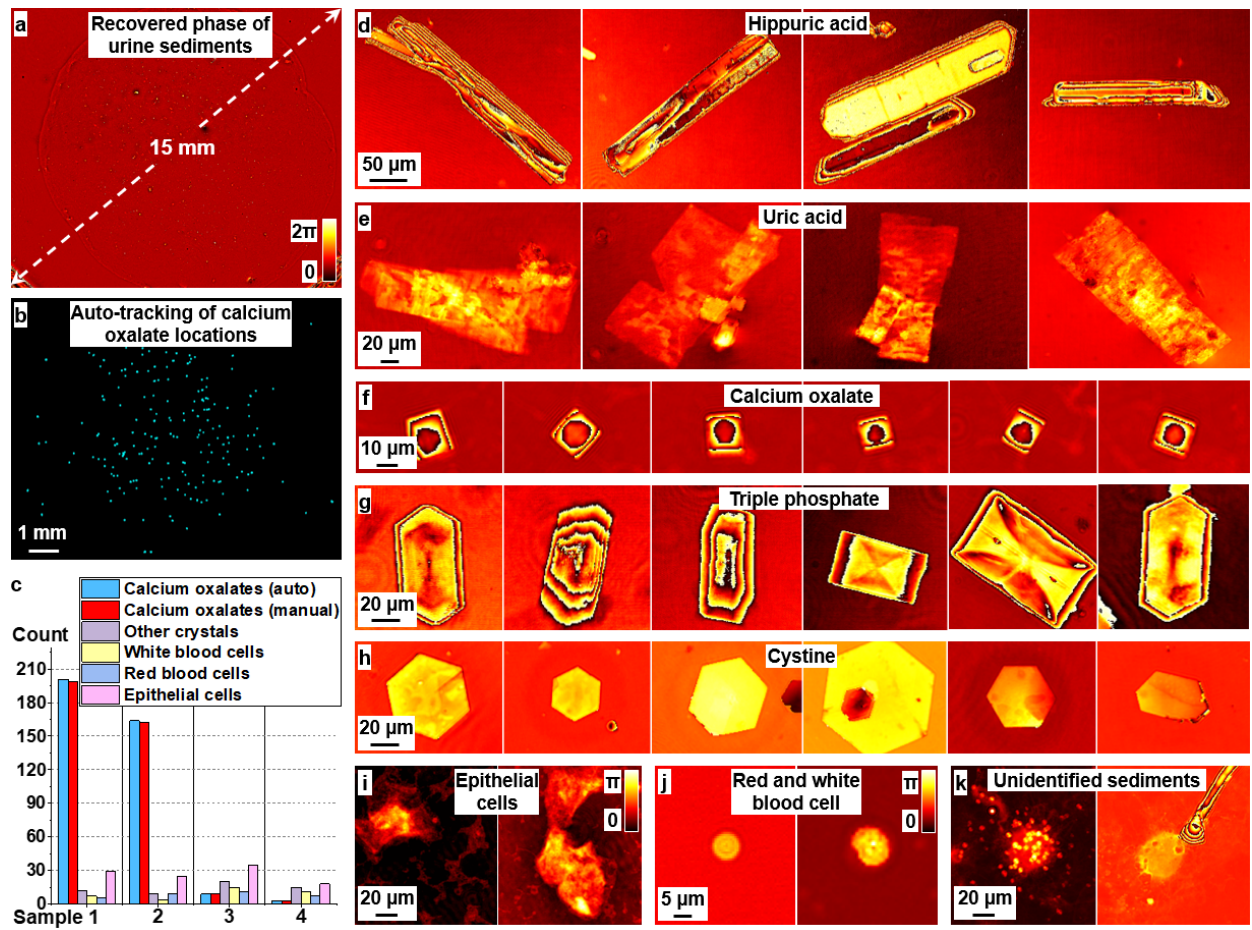

**Fig. 4. High-throughput automatic urine sediment examination.** (a) The recovered whole slide phase image of a urine sediment sample. (b) The automatic tracking of calcium oxalate crystals over the entire field of view. (c) The counting of different types of crystals and cells of 4 urine samples. (d-k) The recovered phase images of different elements in the urine sediment slides, including hippuric acid crystals (d), uric acid crystals (e), calcium oxalate crystals (f), triple phosphate crystals (g), cystine crystals (h), epithelial cells (i), red and white blood cells (j), and other unidentified sediments (k). Figure S4 shows sample images of calcium oxalate crystals miscounted by the automatic tracking algorithm. Other sample images of unidentified sediments are shown in Figure S6.

**High-throughput cytometric analysis of cells.** Based on the recovered gigapixel image of the reported device, we can perform automatic cell segmentation and counting for high-throughput cytometric analysis. Figure 5 demonstrates the use of the recovered phase for automatically tracking white blood cells (WBCs) and *Trypanosoma brucei* parasites in a blood smear sample. The inset of Figure 5a shows the locations of WBCs and parasites, marked with red and blue dots, respectively. In Figure 5b, we analyzed the cell area and the average phase for the WBCs and the *Trypanosomes*. We can see that the scattering plot shows two clusters that correspond to these two types of cells (we have manually examined ~150 cells at the boundary of these two clusters for proper color labeling in Figure 5b). Figure 5c shows the magnified view of the region in Figure 5a. Figures 5d1 and 5d2 show the recovered phase and intensity images of a small region in Figure 5c. As a comparison, we also show the intensity image captured with a 20×, 0.75 NA objective lens in Figure 5d3.

In Figure S7, we further demonstrate the automatic segmentation of WBCs, red blood cells (RBCs), and parasites based on the average phase and cell area of the recovered phase image. Figure S7a and S7b show the recovered images of the region in Figure 5c. Figure S7c shows the results of automatic cell segmentation and counting of the same region. The results of our automatic counting and manual counting using a regular light microscope are shown in Figure S7d, where they are in good agreement with each other for this region. For the entire field of view shown in Figure 5a, we have identified ~850,000 cells, and the acquisition time is ~90 seconds for covering 6 fields of view of the blood-coated sensor. The corresponding acquisition throughput is ~10000 cells per second, which is comparable to a regular flow cytometer while with a small fraction of the cost. A regular non-imaging flow cytometer, on the other hand, cannot render microscopic images and the sample cannot be used for further analysis at a later time point. We note that there are optofluidic platforms that can image cells with much higher throughput (52, 53). For example, Holzner *et. al.* demonstrated the use of a microlens array for imaging cells at a throughput of 50000 cells per second. The key advantage of these optofluidic platforms is the efficient image acquisition with the help of hydrodynamic focusing. The reported device, on the other hand, has the flexibility to image different types of specimens, from regular pathology slides, Petri dishes, to 96 well plates. The low-cost nature also allows us to scale up the throughput by simply using multiple image sensors (45).

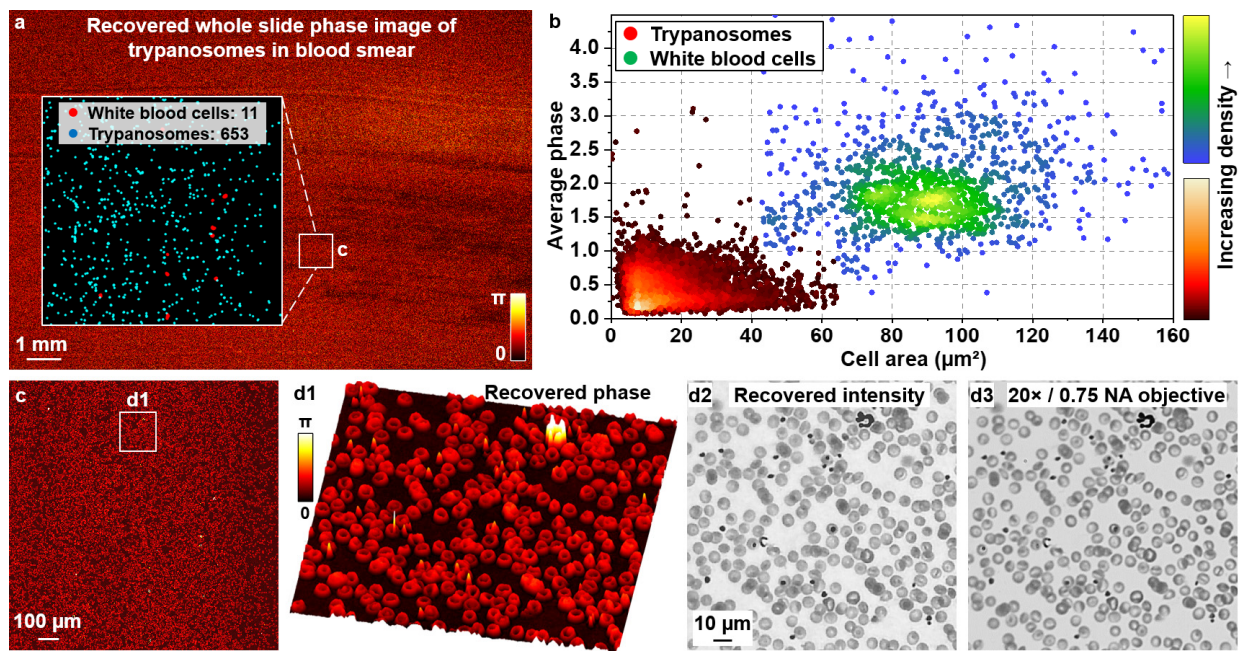

**Fig. 5. High-throughput cytometric analysis of trypanosomes and blood cells.** (a) The recovered whole slide phase image of a blood smear with *Trypanosoma* parasites. Inset shows the locations of WBCs and parasites based on the automatic segmentation and tracking process. (b) The scattering plot of WBCs and parasites based on cell area and average phase. (c) The magnified view of the small region in (a). The recovered phase (d1) and intensity (d2) of the small region in (c). (d3) The image captured using a regular light microscope with a 20×, 0.75 NA Nikon objective lens.

***In vitro* monitoring of bacterial culture over the entire Petri dish.** The reported device can also be used to monitor longitudinal cell culture experiments within an incubator. The combination of disc rotation and sensor translation allows us to cover the area of the entire Petri dish. In our experiment, 10 mL fresh

Mueller-Hinton broth was first inoculated with a single colony of *E. coli* ATCC 25922 strain from the Mueller-Hinton agar plate and kept at 37 °C. The culture was then incubated in a culture tube at 37 °C overnight. On the following day, we adjusted the turbidity of the bacterial solution to 0.5 McFarland standard with fresh Mueller-Hinton medium, containing ~$10^8$ CFU / mL. The bacteria suspension was then diluted to a concentration of ~$10^3$ CFU / mL. The concentration of the final diluted bacteria suspension was also checked using the standard plate count method. We then added the prepared bacteria suspension to a regular Petri dish with Mueller-Hinton agar for imaging. Since bacterial growth can be prohibited by violet light, we used the 650 nm laser diode from the Blu-ray drive in this experiment.

Figure 6 shows the recovered phase image of live *E coli*. bacterial cultures on the uneven agar plate. Figure 6a shows the entire field of view with ~35 mm in diagonal. The image of Figure 6a contains ~2 gigapixels and it took 5-6 mins to acquire the entire dataset. Figures 6b1-6b8 show the magnified views of the different regions in Figure 6a. The top panels of Figure 6b are the recovered phase using the reported device. The bottom panels show the unwrapped phase of the colonies and they can be used to quantify the dry mass and quantitatively track the bacterial growth (48). With the single cell imaging result in Figure 3d, we can see that the reported device can bridge the scale between a single bacterial cell to the entire Petri dish platform. This unique capability can be used to monitor spatiotemporal microbial evolution on antibiotic landscapes. Figure 6b shows the phase images of colonies at different locations. In Fig. S8, we show the same colony acquired at different time points.

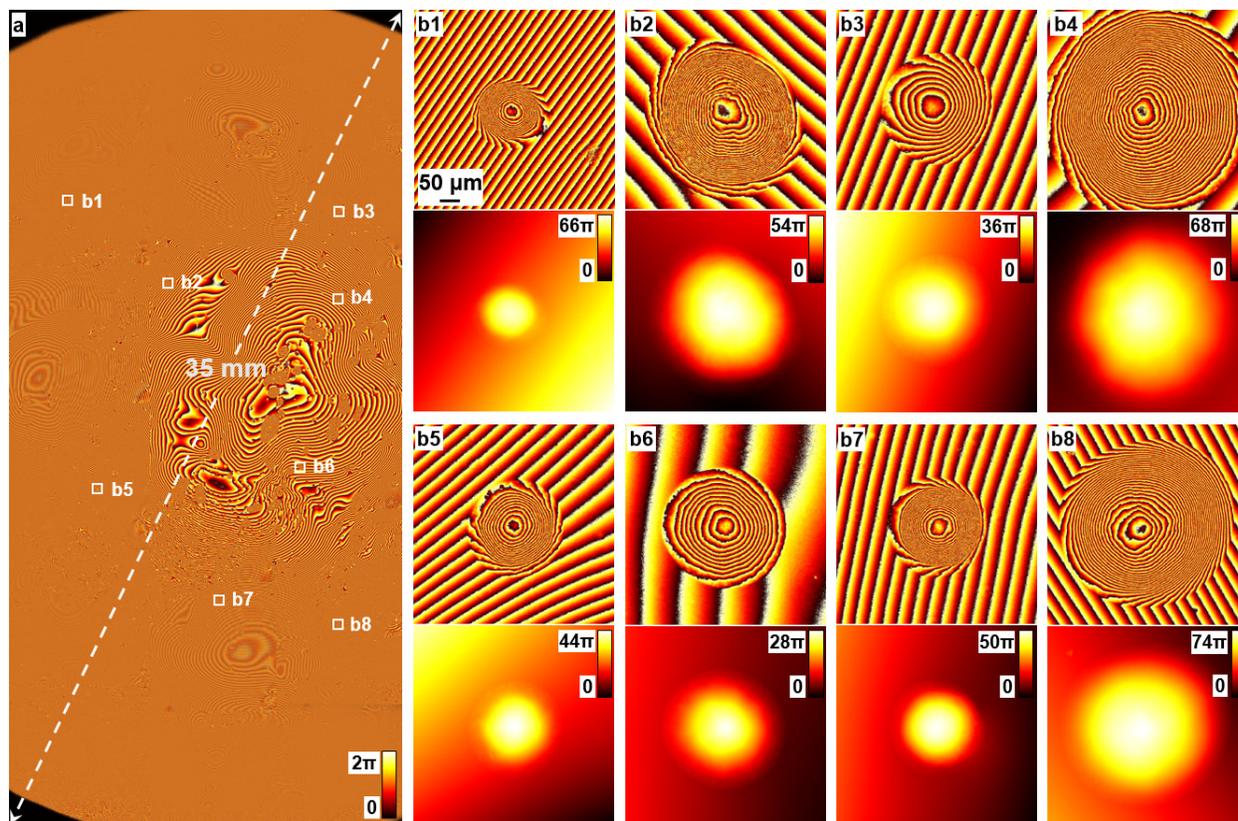

**Fig. 6. Imaging the entire Petri dish for *in vitro* monitoring of bacterial cultures.** (a) The recovered phase image of the entire Petri dish with ~35 mm in diagonal. (b1-b8) The magnified views of different regions of the Petri dish, where bacterial colonies are forming on the uneven plate. The top panels show the recovered phase images using the reported device. The bottom panels show the unwrapped phase images that can be used to measure the growth of bacteria over time. Refer to Figure S8 for the colony growth at different time points.

**Discussion and conclusion**

In summary, we have reported the integration of a blood-coated image sensor with a modified Blu-ray drive for large-scale, high-resolution microscopic imaging. Based on the disc spinning process of the reported device, we have also developed and implemented a new strategy for diversity measurement, termed rotational ptychography. In our device, the rich spatial feature of the blood-cell monolayer can down-modulate the diffracted waves for sensor detection, thus forming a high-resolution computational bio-lens. In our characterization experiment, we can resolve the 435 nm linewidth on the resolution target. We can also clearly resolve individual grape-like cells of Staphylococcus aureus bacteria. Compared to the regular optical lens, this blood-cell monolayer has a theoretically unlimited field of view, thereby addressing the trade-off between the resolution and field of view in conventional microscope platforms.

To demonstrate the imaging capabilities of the reported device, we performed high-throughput urinalysis by locating disease-related crystals over a centimeter-scale area. The recovered phase of these crystals contains many $2\pi$ wraps. We note that it is challenging for other common multi-height and multi-wavelength lensless techniques to properly restore the phase wraps of complex objects. For cytometric analysis, we quantified the labeled cells in a blood film at an acquisition speed of ~10,000 cells per second, achieving an acquisition throughput similar to that of a regular flow cytometer but at a small fraction of the cost. For *in vitro* demonstration, we monitored live bacterial cultures over the entire Petri dish of ~35 mm in diameter. The capability of bridging the scale between a single bacterial cell to the entire Petri dish platform allows the monitoring of spatiotemporal microbial evolution on antibiotic landscapes.

Measurements and statistical analysis of structural, morphological, and chemical phenotypes of cells and bio-objects using the reported device are essential for providing insights into biomedical processes. Obtaining high-content images of bio-specimens using the reported device is also well aligned with the pressing need for large-scale datasets for machine learning to make better decisions in clinical diagnosis. The future applications of the reported device include digital pathology, microbiology, hematology, drug screening, and various point-of-care diagnostics. The quantitative phase imaging capability also enables a wide range of biomedicine-related applications. For example, the recovered phase image can be converted into a quantitative height map for 3D visualization. It can be used to quantify the growth of bacterial micro-colonies. Lastly, it is the first demonstration of a large-scale computational bio-lens for high-throughput CDI. Other parts of the device, including the coherent light source, can be obtained from a Blu-ray drive. With the reported device, we envision the widespread of high-throughput optical microscopy from well-equipped laboratories to citizen scientists worldwide.

**Supporting Information**

Video 1: operation of the reported device. Figures S1-S8: wiring diagram for controlling the laser diodes in the Blu-ray drive, initial positional shift estimation in rotational ptychography, recovery of the disc motion based on the initial positional shift, sample images of calcium oxalate crystals miscounted by the automatic tracking algorithm, sample images of unidentified sediments in urine samples, the recovered images of urine crystals and the comparison with those captured with the regular light microscope, automatic segmentation and counting of white blood cells, red blood cells, and trypanosomes, detection of bacterial growth at different time points. Note 1: motion tracking for rotational ptychography.

**Author Contributions**

S. J., C. G., and T. W. developed the prototype system and prepared the display items. J. L, and P. S. prepared and analysed the data for rapid urine sediment imaging. G. Z. conceived the idea and supervised the project. All authors contributed to manuscript preparation and editing.

**Notes**

The authors declare that they have no known competing financial interests or personal relationships that could have appeared to influence the work reported in this paper.

**Funding**

This work was partially supported by National Science Foundation 1700941, 2012140 and the UConn SPARK grant. P. S. acknowledges the support of the Thermo Fisher Scientific fellowship.

**Reference:**

(1) Lohmann, A. W.; Dorsch, R. G.; Mendlovic, D.; Zalevsky, Z.; Ferreira, C., Space–bandwidth product of optical signals and systems. *JOSA A* **1996,** *13* (3), 470-473.

(2) Park, J.; Brady, D. J.; Zheng, G.; Tian, L.; Gao, L., Review of bio-optical imaging systems with a high space-bandwidth product. *Advanced Photonics* **2021,** *3* (4), 044001.

(3) Zheng, G.; Shen, C.; Jiang, S.; Song, P.; Yang, C., Concept, implementations and applications of Fourier ptychography. *Nature Reviews Physics* **2021,** *3* (3), 207-223.

(4) Zheng, G.; Horstmeyer, R.; Yang, C., Wide-field, high-resolution Fourier ptychographic microscopy. *Nature photonics* **2013,** *7* (9), 739.

(5) Dong, S.; Horstmeyer, R.; Shiradkar, R.; Guo, K.; Ou, X.; Bian, Z.; Xin, H.; Zheng, G., Aperture-scanning Fourier ptychography for 3D refocusing and super-resolution macroscopic imaging. *Optics express* **2014,** *22* (11), 13586-13599.

(6) Zuo, C.; Sun, J.; Li, J.; Asundi, A.; Chen, Q., Wide-field high-resolution 3D microscopy with Fourier ptychographic diffraction tomography. *Optics and Lasers in Engineering* **2020,** *128*, 106003.

(7) McConnell, G.; Trägårdh, J.; Amor, R.; Dempster, J.; Reid, E.; Amos, W. B., A novel optical microscope for imaging large embryos and tissue volumes with sub-cellular resolution throughout. *Elife* **2016,** *5*, e18659.

(8) Fan, J.; Suo, J.; Wu, J.; Xie, H.; Shen, Y.; Chen, F.; Wang, G.; Cao, L.; Jin, G.; He, Q., Video-rate imaging of biological dynamics at centimetre scale and micrometre resolution. *Nature Photonics* **2019,** *13* (11), 809-816.

(9) Greenbaum, A.; Luo, W.; Su, T.-W.; Göröcs, Z.; Xue, L.; Isikman, S. O.; Coskun, A. F.; Mudanyali, O.; Ozcan, A., Imaging without lenses: achievements and remaining challenges of wide-field on-chip microscopy. *Nature methods* **2012,** *9* (9), 889-895.

(10) Fienup, J. R., Phase retrieval algorithms: a comparison. *Applied optics* **1982,** *21* (15), 2758-2769.

(11) Bian, Z.; Guo, C.; Jiang, S.; Zhu, J.; Wang, R.; Song, P.; Zhang, Z.; Hoshino, K.; Zheng, G., Autofocusing technologies for whole slide imaging and automated microscopy. *Journal of Biophotonics* **2020,** *13* (12), e202000227.

(12) Hwu, E. E.-T.; Boisen, A., Hacking CD/DVD/Blu-ray for biosensing. *ACS sensors* **2018,** *3* (7), 1222-1232.

(13) Nolte, D. D.; Regnier, F. E., Spinning-disk interferometry: The biocd. *Optics and photonics news* **2004,** *15* (10), 48-53.

(14) Tseng, D.; Mudanyali, O.; Oztoprak, C.; Isikman, S. O.; Sencan, I.; Yaglidere, O.; Ozcan, A., Lensfree microscopy on a cellphone. *Lab on a Chip* **2010,** *10* (14), 1787-1792.

(15) Li, P.; Maiden, A., Lensless LED matrix ptychographic microscope: problems and solutions. *Applied optics* **2018,** *57* (8), 1800-1806.

(16) Zhang, Z.; Zhou, Y.; Jiang, S.; Guo, K.; Hoshino, K.; Zhong, J.; Suo, J.; Dai, Q.; Zheng, G., Invited Article: Mask-modulated lensless imaging with multi-angle illuminations. *APL Photonics* **2018,** *3* (6), 060803.

(17) Zhou, Y.; Wu, J.; Suo, J.; Han, X.; Zheng, G.; Dai, Q., Single-shot lensless imaging via simultaneous multi-angle LED illumination. *Optics express* **2018,** *26* (17), 21418-21432.
(18) Zheng, G.; Lee, S. A.; Antebi, Y.; Elowitz, M. B.; Yang, C., The ePetri dish, an on-chip cell imaging platform based on subpixel perspective sweeping microscopy (SPSM). *Proceedings of the National Academy of Sciences* **2011,** *108* (41), 16889-16894.
(19) Wang, X.; Zhao, M.; Nolte, D. D., Prostate-specific antigen immunoassays on the BioCD. *Analytical and bioanalytical chemistry* **2009,** *393* (4), 1151-1156.
(20) Lee, B. S.; Lee, Y. U.; Kim, H.-S.; Kim, T.-H.; Park, J.; Lee, J.-G.; Kim, J.; Kim, H.; Lee, W. G.; Cho, Y.-K., Fully integrated lab-on-a-disc for simultaneous analysis of biochemistry and immunoassay from whole blood. *Lab on a Chip* **2011,** *11* (1), 70-78.
(21) Kim, T.-H.; Park, J.; Kim, C.-J.; Cho, Y.-K., Fully integrated lab-on-a-disc for nucleic acid analysis of food-borne pathogens. *Analytical chemistry* **2014,** *86* (8), 3841-3848.
(22) Miccio, L.; Memmolo, P.; Merola, F.; Netti, P.; Ferraro, P., Red blood cell as an adaptive optofluidic microlens. *Nature communications* **2015,** *6* (1), 1-7.
(23) De Tommasi, E.; De Luca, A.; Lavanga, L.; Dardano, P.; De Stefano, M.; De Stefano, L.; Langella, C.; Rendina, I.; Dholakia, K.; Mazilu, M., Biologically enabled sub-diffractive focusing. *Optics express* **2014,** *22* (22), 27214-27227.
(24) Zhang, H.; Jiang, S.; Liao, J.; Deng, J.; Liu, J.; Zhang, Y.; Zheng, G., Near-field Fourier ptychography: super-resolution phase retrieval via speckle illumination. *Optics express* **2019,** *27* (5), 7498-7512.
(25) Zhang, H.; Bian, Z.; Jiang, S.; Liu, J.; Song, P.; Zheng, G., Field-portable quantitative lensless microscopy based on translated speckle illumination and sub-sampled ptychographic phase retrieval. *Optics letters* **2019,** *44* (8), 1976-1979.
(26) Song, P.; Jiang, S.; Zhang, H.; Bian, Z.; Guo, C.; Hoshino, K.; Zheng, G., Super-resolution microscopy via ptychographic structured modulation of a diffuser. *Optics letters* **2019,** *44* (15), 3645-3648.
(27) Jiang, S.; Zhu, J.; Song, P.; Guo, C.; Bian, Z.; Wang, R.; Huang, Y.; Wang, S.; Zhang, H.; Zheng, G., Wide-field, high-resolution lensless on-chip microscopy via near-field blind ptychographic modulation. *Lab on a Chip* **2020,** *20* (6), 1058-1065.
(28) Jang, M.; Horie, Y.; Shibukawa, A.; Brake, J.; Liu, Y.; Kamali, S. M.; Arbabi, A.; Ruan, H.; Faraon, A.; Yang, C., Wavefront shaping with disorder-engineered metasurfaces. *Nature photonics* **2018,** *12* (2), 84-90.
(29) Choi, Y.; Yoon, C.; Kim, M.; Choi, W.; Choi, W., Optical imaging with the use of a scattering lens. *IEEE Journal of Selected Topics in Quantum Electronics* **2013,** *20* (2), 61-73.
(30) Kim, M.; Choi, W.; Choi, Y.; Yoon, C.; Choi, W., Transmission matrix of a scattering medium and its applications in biophotonics. *Optics express* **2015,** *23* (10), 12648-12668.
(31) Wu, X.; Sun, J.; Zhang, J.; Lu, L.; Chen, R.; Chen, Q.; Zuo, C., Wavelength-scanning lensfree on-chip microscopy for wide-field pixel-super-resolved quantitative phase imaging. *Optics Letters* **2021,** *46* (9), 2023-2026.
(32) Luo, W.; Zhang, Y.; Feizi, A.; Göröcs, Z.; Ozcan, A., Pixel super-resolution using wavelength scanning. *Light: Science & Applications* **2016,** *5* (4), e16060-e16060.
(33) Dean, B. H.; Bowers, C. W., Diversity selection for phase-diverse phase retrieval. *JOSA A* **2003,** *20* (8), 1490-1504.
(34) Guizar-Sicairos, M.; Fienup, J. R., Phase retrieval with transverse translation diversity: a nonlinear optimization approach. *Optics express* **2008,** *16* (10), 7264-7278.
(35) Zhang, Y.; Pedrini, G.; Osten, W.; Tiziani, H. J., Whole optical wave field reconstruction from double or multi in-line holograms by phase retrieval algorithm. *Optics Express* **2003,** *11* (24), 3234-3241.
(36) Guo, C.; Jiang, S.; Song, P.; Wang, T.; Shao, X.; Zhang, Z.; Zheng, G., Quantitative multi-height phase retrieval via a coded image sensor. *Biomedical Optics Express* **2021,** *12* (11), 7173-7184.
(37) Jiang, S.; Guo, C.; Hu, P.; Hu, D.; Song, P.; Wang, T.; Bian, Z.; Zhang, Z.; Zheng, G., High-throughput lensless whole slide imaging via continuous height-varying modulation of a tilted sensor. *Optics Letters* **2021,** *46* (20), 5212-5215.

(38) Bao, P.; Zhang, F.; Pedrini, G.; Osten, W., Phase retrieval using multiple illumination wavelengths. *Optics letters* **2008,** *33* (4), 309-311.
(39) Faulkner, H. M. L.; Rodenburg, J., Movable aperture lensless transmission microscopy: a novel phase retrieval algorithm. *Physical review letters* **2004,** *93* (2), 023903.
(40) Dierolf, M.; Menzel, A.; Thibault, P.; Schneider, P.; Kewish, C. M.; Wepf, R.; Bunk, O.; Pfeiffer, F., Ptychographic X-ray computed tomography at the nanoscale. *Nature* **2010,** *467* (7314), 436-439.
(41) Thibault, P.; Menzel, A., Reconstructing state mixtures from diffraction measurements. *Nature* **2013,** *494* (7435), 68-71.
(42) Maiden, A. M.; Rodenburg, J. M., An improved ptychographical phase retrieval algorithm for diffractive imaging. *Ultramicroscopy* **2009,** *109* (10), 1256-1262.
(43) Thibault, P.; Dierolf, M.; Bunk, O.; Menzel, A.; Pfeiffer, F., Probe retrieval in ptychographic coherent diffractive imaging. *Ultramicroscopy* **2009,** *109* (4), 338-343.
(44) Rodenburg, J.; Maiden, A., Ptychography. In *Springer Handbook of Microscopy*, Springer: 2019; pp 819-904.
(45) Jiang, S.; Guo, C.; Song, P.; Zhou, N.; Bian, Z.; Zhu, J.; Wang, R.; Dong, P.; Zhang, Z.; Liao, J.; Yao, J.; Feng, B.; Murphy, M.; Zheng, G., Resolution-Enhanced Parallel Coded Ptychography for High-Throughput Optical Imaging. *ACS Photonics* **2021,** *8* (11), 3261-3271.
(46) Stockmar, M.; Cloetens, P.; Zanette, I.; Enders, B.; Dierolf, M.; Pfeiffer, F.; Thibault, P., Near-field ptychography: phase retrieval for inline holography using a structured illumination. *Scientific reports* **2013,** *3* (1), 1-6.
(47) Song, P.; Guo, C.; Jiang, S.; Wang, T.; Hu, P.; Hu, D.; Zhang, Z.; Feng, B.; Zheng, G., Optofluidic ptychography on a chip. *Lab on a Chip* **2021,** *21* (23), 4549-4556.
(48) Jiang, S.; Guo, C.; Bian, Z.; Wang, R.; Zhu, J.; Song, P.; Hu, P.; Hu, D.; Zhang, Z.; Hoshino, K.; Feng, B.; Zheng, G., Ptychographic sensor for large-scale lensless microbial monitoring with high spatiotemporal resolution. *Biosensors and Bioelectronics* **2022,** *196*, 113699.
(49) Hoppe, W., Diffraction in inhomogeneous primary wave fields. 1. Principle of phase determination from electron diffraction interference. *Acta Crystallographica Section a-Crystal Physics Diffraction Theoretical and General Crystallography* **1969**, 495-&.
(50) Bian, Z.; Jiang, S.; Song, P.; Zhang, H.; Hoveida, P.; Hoshino, K.; Zheng, G., Ptychographic modulation engine: a low-cost DIY microscope add-on for coherent super-resolution imaging. *Journal of Physics D: Applied Physics* **2019,** *53* (1), 014005.
(51) Maiden, A.; Johnson, D.; Li, P., Further improvements to the ptychographical iterative engine. *Optica* **2017,** *4* (7), 736-745.
(52) Holzner, G.; Du, Y.; Cao, X.; Choo, J.; deMello, A. J.; Stavrakis, S., An optofluidic system with integrated microlens arrays for parallel imaging flow cytometry. *Lab on a Chip* **2018,** *18* (23), 3631-3637.
(53) Lei, C.; Kobayashi, H.; Wu, Y.; Li, M.; Isozaki, A.; Yasumoto, A.; Mikami, H.; Ito, T.; Nitta, N.; Sugimura, T., High-throughput imaging flow cytometry by optofluidic time-stretch microscopy. *Nature protocols* **2018,** *13* (7), 1603-1631.

For Table of Contents Use Only

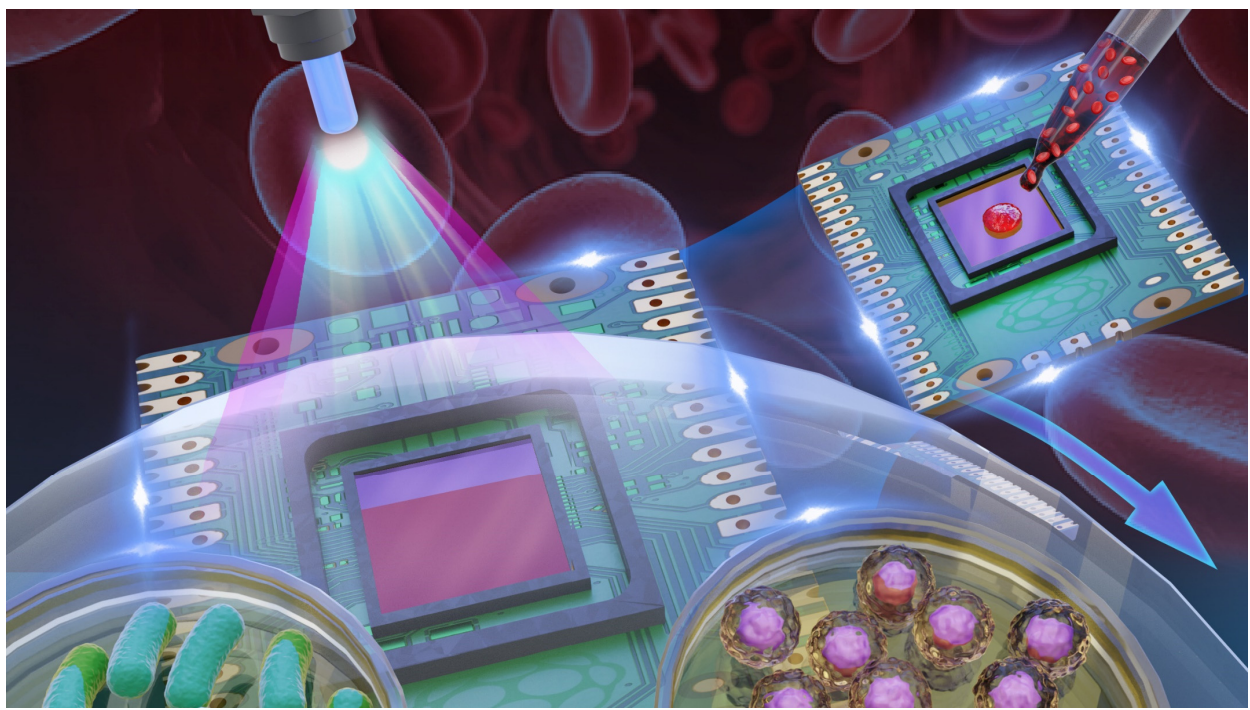

This work integrates a blood-coated image sensor with a modified Blu-ray drive for high-throughput lensless microscopy. The built-in lasers of the Blu-ray drive are used to illuminate the specimens mounted on the spinning disc. The resulting diffraction patterns are recorded using the blood-coated image sensor for ptychographic reconstruction. The device can resolve the 435-nm linewidth on the resolution target and has an imaging throughput of ~10,000 cells per second. Live cell cultures can be monitored over the entire Petri dish with high resolution.